\documentclass[showpacs,preprintnumbers,amsmath,prl,
graphicx,amssymb,superscriptaddress,twocolumn,nofootinbib]{revtex4}
\usepackage{graphicx}
\usepackage{dcolumn}
\usepackage{bm}

\renewcommand{\today}{\number\day\space\ifcase\month\or January\or 
 February\or March\or April\or May\or June\or July\or August\or 
 September\or October\or November\or December\fi\space\number\year}

\begin{document}

\title{Comments on "A Dark Matter Search with MALBEK"}

%%%%%%%%%%%%

\newcommand{\efi}{Enrico Fermi Institute, Kavli Institute for 
Cosmological Physics and Department of Physics,
University of Chicago, Chicago, IL 60637}
%%%%%%%%%%%%

\affiliation{\efi}
														
\author{J.I.~Collar}\affiliation{\efi}

\begin{abstract}

CoGeNT and MALBEK use p-type point contact germanium detectors to search for low-mass dark matter particles. Both detectors enjoy identical intrinsic noise characteristics. However, MALBEK's data acquisition electronics severely degrade the ability to separate signals originating in the bulk of the germanium crystal from surface backgrounds, through a measurement of the preamplifier pulse rise-time in the sub-keVee energy range of interest. The physical meaning of the parameter W$_{par}$ developed by the MAJORANA collaboration to compensate for this limitation is clarified here. It is shown that this parameter does not correlate to rise-time at low energy, and is presently unable to distinguish between surface and bulk events below $\sim$1 keVee. This leads to a sizable overstatement of MALBEK's sensitivity to low-mass dark matter particles, when employing aggressive W$_{par}$ cuts.

\end{abstract}

%% ENTER PACS NEXT
%\pacs{85.30.-z, 95.35.+d, 95.55.Vj, 14.80.Mz}

\maketitle

%%%%%%%%%%%%%%%%%%%%%%%%%%%%%%%%%%%%%%%%%%%%%%%%%%%%%%%%%%%%%%%%%%%%%%

This brief note intends to clarify the important differences in performance at low-energy between the CoGeNT and MALBEK detectors. Strong limitations in both hardware and methodology affecting MALBEK are hard to derive from the discussion offered in \cite{malbek}. Both detectors are very similar in their intrinsic characteristics (noise, background, capacitance, design, etc.\ \cite{paddy}), with low-background cryostat parts and inner shielding for MALBEK being provided by the CoGeNT collaboration. The MALBEK detector is by design a replica of the CoGeNT detector, and as such it displays the same intrinsic electronic noise ($\sim$160 eV FWHM). However, the MALBEK data acquisition system, intended for future use in the MAJORANA Demonstrator array, severely curtails this performance by injecting a dominant level of electronic noise (Fig.\ 1). This problem persisted even following attempts to ameliorate the polling noise from a Struck SIS3302 VME digitizer through software modifications \cite{paddy}.

\begin{figure}[!htbp]
\includegraphics[width=7.cm]{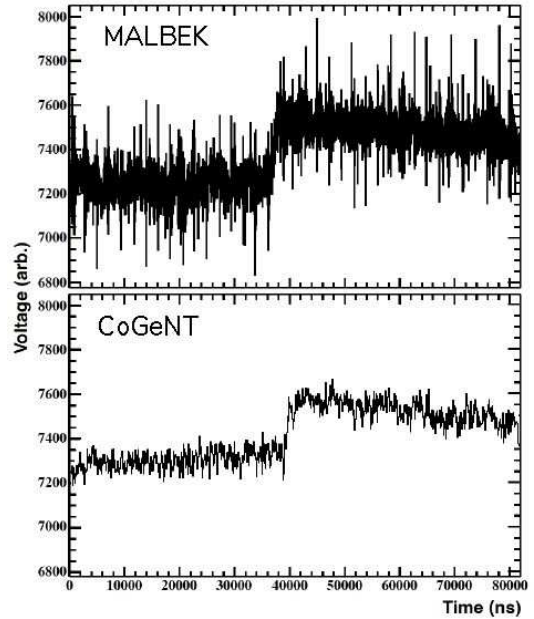}
\caption{Direct comparison between MALBEK  and CoGeNT preamplifier traces corresponding to typical 1.19 keVee energy depositions, both prior to any denoising. MALBEK's trace is from  \protect\cite{paddy}. The sampling period in MALBEK is 10 ns, and 50 ns for CoGeNT. Both detectors enjoy the same intrinsic electronic noise ($\sim$160 eV FWHM). The additional injection of noise by MALBEK's data acquisition system, evident in this figure, severely limits the ability to correctly identify pulse rise-times below $\sim$4 keVee \protect\cite{paddy}. This rise-time is the physical property employed to separate bulk signals from surface backgrounds in PPC detectors, being inversely proportional to the electric field in these two detector regions, the magnitude of which affects charge mobility \protect\cite{Aal11,Aal11b,longcogent}.}
\end{figure}

As is acknowledged in \cite{paddy}, this additional source of electronic noise makes it impossible to adequately measure the rise-time (t$_{10-90}$) of preamplifier pulses below $\sim$4 keVee in MALBEK  (Fig.\ 2, top panel). This rise-time is the parameter that enables the separation between bulk (small  t$_{10-90}$) signals like those expected from hypothetical dark matter particles, and surface (large  t$_{10-90}$) backgrounds within CoGeNT. Rise-time is physically correlated to the intensity of the electric field in these two regions, through its effect on charge mobility \cite{Aal11,Aal11b,longcogent}. To compensate for this limitation, MALBEK adopts a parameter (W$_{par}$) derived from wavelet analysis of preamplifier traces, to attempt a discrimination between surface and bulk events. This parameter is formally defined as W$_{par}\!=\!$ max$(\mid\!\!c_{D}^{(i)}(n=0)\!\!\mid ^{2})/E^{2}$, or roughly equivalent to a "smoothed out derivative normalized by the square of the energy ($E$) of the event" \cite{paddy}. The coefficients $c_{D}^{(i)}(n=0)$ are the so-called detail coefficients corresponding to the  first iteration $(n=0)$ of a stationary wavelet transform (SWT), where the index $(i)$ corresponds to each sampled datapoint in the preamplifier pulse trace. This {\it prima facie} rather obscure definition of W$_{par}$ will be further clarified below. W$_{par}$ and the rise-time of preamplifier traces are observed to correlate in MALBEK for calibration events induced by a low-energy $^{241}$Am source, capable of producing both fast (bulk) and slow (surface) events. However, this strongly non-linear correlation is demonstrated only above 4 keVee, due to the impossibility to correctly reconstruct rise-times at lower energies \cite{paddy} (Fig.\ 2, top panel). In addition to this $^{241}$Am calibration, surface events arising from the presence of a contemporary lead patch during initial MALBEK runs were observed to accumulate in the broad range 1000 $<$W$_{par}<$ 3000 next to the 0.6 keVee MALBEK threshold \cite{paddy}, providing a useful reference. Electronic pulser events with a fixed rise-time, able to approximate fast bulk events, were also used to characterize the dependence of W$_{par}$ on energy (Fig.\ 2, bottom panel). 

\begin{figure}
\includegraphics[width=7.cm]{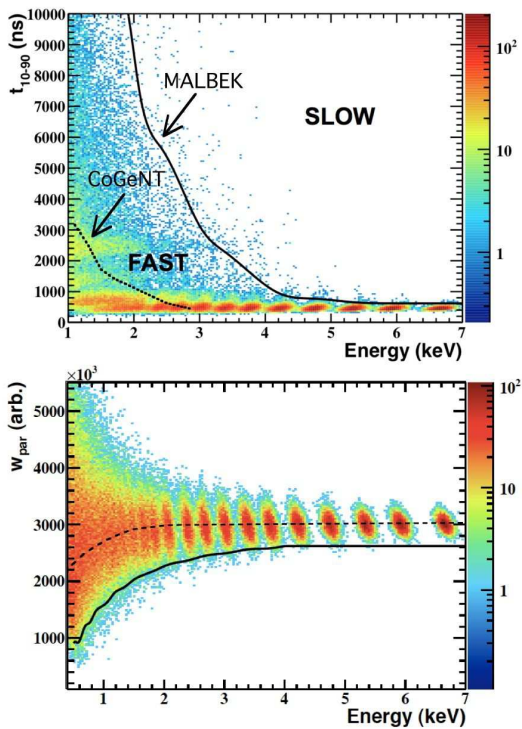}
\caption{(adapted from \protect\cite{paddy}) {\it Top:} 99\% acceptance boundary for the measured rise-time of fast electronic pulser signals (approximate replicas of bulk events) in MALBEK. The noise injected by MALBEK's electronics corrupts this rise-time measurement below few keVee \protect\cite{paddy}. Overlapped on the figure is the same 99\% boundary for CoGeNT pulser runs, reaching t$_{10-90}\sim\!7\mu$s at 0.5 keVee, CoGeNT's analysis threshold. {\it Bottom:} distribution of W$_{par}$ and 99\% acceptance boundary (solid line) for electronic pulser events in MALBEK. MALBEK uses this alternative parameter in an attempt to perform bulk/surface discrimination (see text). Overlapped on the figure (dotted line) is the mean of W$_{par}$ as a function of energy, estimated and added by this author through an analysis of the color scale. A discussion of this migration  of W$_{par}$ towards lower values with decreasing energy, also noticeable in MALBEK's physics runs (Fig.\ 3), is absent from both \protect\cite{malbek} and  \protect\cite{paddy}. This migration results into an almost complete overlap of the W$_{par}$ distributions for surface and bulk events at 0.6 keVee, MALBEK's analysis threshold (see text).}
\end{figure}

Unfortunately, a discussion of several crucial limitations to the use of W$_{par}$ to distinguish between bulk and surface events is absent from both \cite{malbek} and \cite{paddy}. In particular, the choice of color scale and energy sampling values in the bottom panel of Fig.\ 2 makes it hard to notice that the mean value of W$_{par}$ for electronic pulser events is rapidly migrating towards lower values with decreasing energy. This results into a near complete merger of surface and bulk event W$_{par}$ distributions at 0.6 keVee, MALBEK's analysis threshold. This migration is also readily noticeable when the evolution in energy of W$_{par}$ distributions for MALBEK physics runs is carefully inspected (Fig.\ 3). None of this important information can be found in \cite{malbek,paddy}. This is in contrast to the behavior of t$_{10-90}$ for CoGeNT events, for which a good bulk/surface separation is maintained all the way down to the sub-keVee energy region (Figs.\ 3 and 5), where a low-significance annual modulation is observed to dominate \cite{neal}. In this situation, the application of an aggressive energy-independent cut accepting only events with W$_{par}\!>\!2500$ (Fig.\ 2, \cite{notation}) as bulk events, as is done in \cite{malbek}, is guaranteed to severely underestimate their number below $\sim$1 keVee. This  leads to an artificially improved sensitivity to low-mass WIMPs. Specifically, in view of the complete lack of correlation between W$_{par}$ and rise-time below $\sim$1 keVee elucidated below, the use of an electronic pulser with a fixed t$_{10-90}=0.4 ~\mu$s to establish a low-energy bulk-signal acceptance boundary for W$_{par}$ should be placed into question \cite{note2}. 

The opacity of the discussion in \cite{malbek,paddy} around the central subject of the correlation or not between W$_{par}$ and rise-time at the lowest energies of interest can lead an attentive reader to wonder what it is that W$_{par}$ really measures. The short answer to this question, developed next, is "not the rise-time". 

\begin{widetext}

\begin{figure}
\includegraphics[width=16cm]{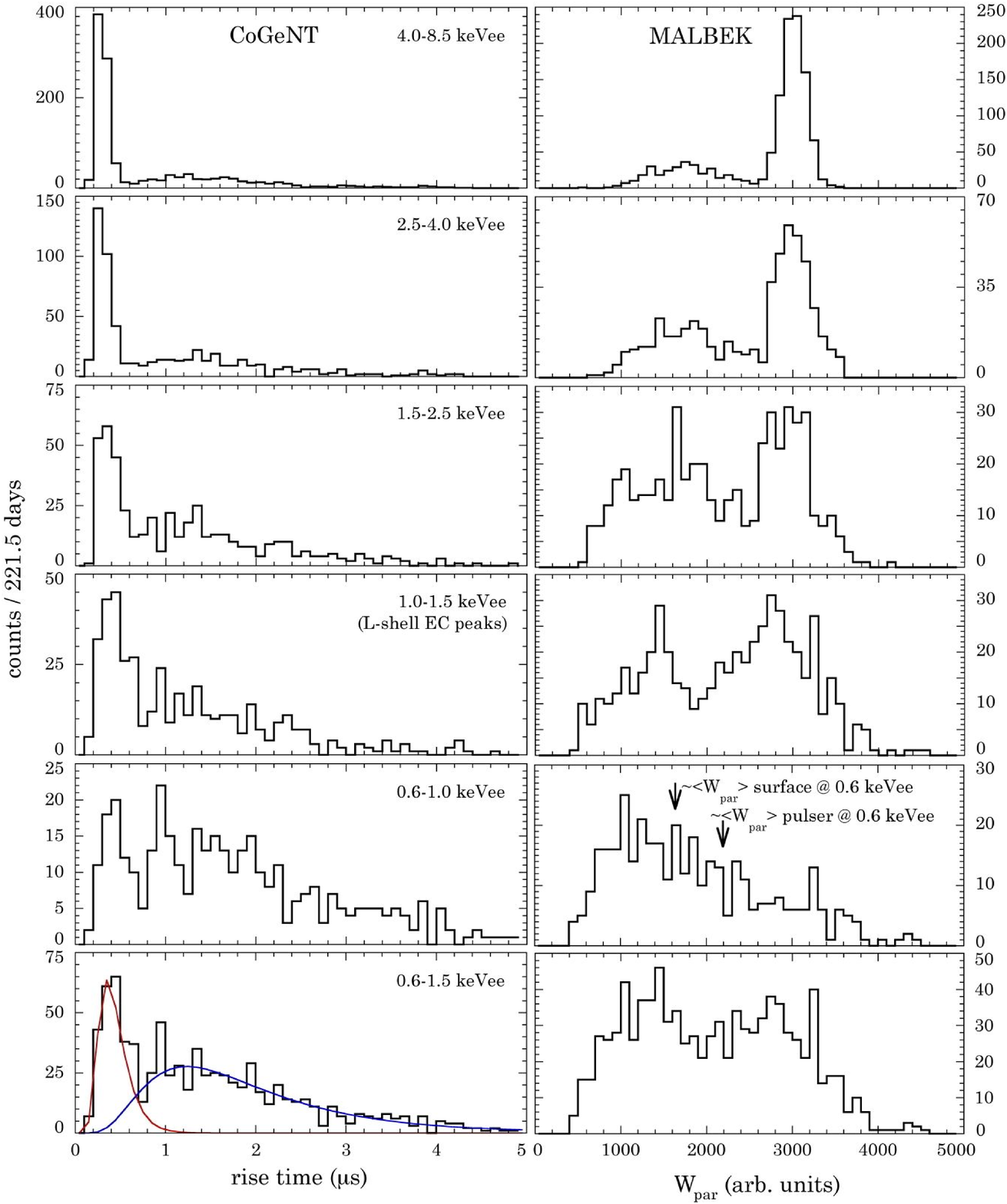}
\caption{Distributions of  t$_{10-90}$ for CoGeNT (left column) and W$_{par}$ in MALBEK (right column), for several energy regions, labelled within left column panels. Bulk pulses are observed as a peak concentrated to the left (right) of CoGeNT (MALBEK) distributions. MALBEK distributions are obtained from the digitization of DS3a and DS3b data (221 days) in \protect\cite{paddy}. The 221 day subset of CoGeNT data shown here was selected to start 637 days from its installation underground, so as to establish a fair comparison to these MALBEK datasets (i.e., a similar level of remaining long-lived cosmogenic activation). A fairly constant separation between bulk events (red in bottom panel) and surface events (blue) is maintained for CoGeNT data all the way down to its 0.5 keVee threshold \protect\cite{longcogent,cogentlast,matt}. In contrast to this, these two distributions are observed to merge together at MALBEK's 0.6 keVee threshold, due to the rapid broadening and migration towards smaller values of  $<$W$_{par}\!>$ for fast events of decreasing energy (see text). Surface MALBEK events were observed to populate the range $\sim1000$ $<$W$_{par}<$ $\sim3000$ next to its threshold, during early runs previous to the removal of internal surface contaminants  \protect\cite{paddy}. }
\end{figure}

\end{widetext}

Taking advantage of the ability to extract rise-time values down to 0.5 keVee in CoGeNT, it is possible to cast light on the opaque meaning of W$_{par}$ by applying both estimators to a recently released CoGeNT dataset  \cite{cogentlast}. The physical meaning of W$_{par}$ can be best understood by observing the behavior of the detail coefficients $c_{D}^{(i)}(n=0)$, on which it depends. Fig.\ 4 shows these coefficients around the position of the preamplifier rising edge ($i\sim6325$), for three well-formed CoGeNT events of 6.8 keVee, large enough  not to be much affected by electronic noise. For sharp pulses with t$_{10-90}\sim0.3 \mu$s (bulk events), a distinct maximum value of $c_{D}^{(i)}(n=0)$, to which W$_{par}$ is proportional by definition, can be easily identified. However, as the rising edge softens (larger t$_{10-90}$, surface events), the maximum absolute $c_{D}^{(i)}(n=0)$ becomes just a random value like that of detail coefficients far from the position of the rising edge in the preamplifier trace. As the energy decreases, pulses are proportionally more affected by electronic noise, and no value of the rise-time, regardless of how small, is able to generate a distinct max$(\mid\!c_{D}^{(i)}(n=0)\!\mid)$ above these random fluctuations (Fig.\ 5). In other words, W$_{par}$ is a simple indicator of how well the SWT is able to reveal the presence of a rising-edge in a preamplifier trace: for large energy pulses well-above the electronic noise of the preamplifier baseline, bulk events return a max$(\mid\!c_{D}^{(i)}(n=0)\!\mid)$ distinct from random fluctuations in these detail coefficients. However, for the present level of electronic noise affecting CoGeNT and MALBEK PPCs, the presence of a rising edge cannot be identified below $\sim$1 keVee through the first iteration of a SWT, regardless of how sharp the transition (how small the rise-time) of a pulse might be: the rising edge is not sufficiently well-defined for the W$_{par}$ estimator to indicate its presence, {\it and all events look like surface events under W$_{par}$ examination}. The behavior displayed in Fig.\ 5 illuminates the origin of the rapid merger of MALBEK's bulk events into the surface event distribution at low energy, already pointed at in Figs.\ 2,3.

\begin{figure}[!htbp]
\includegraphics[width=5cm]{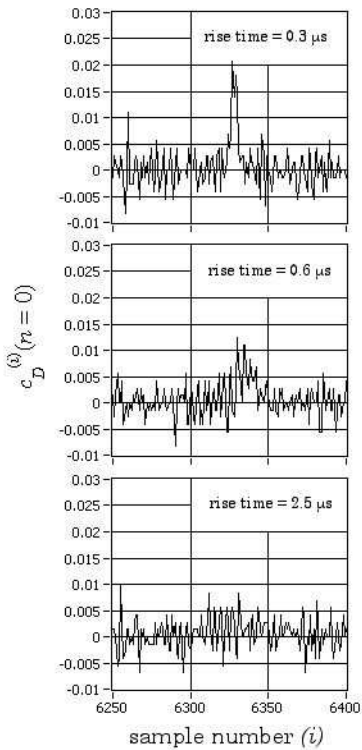}
\caption{Detail coefficients from the first iteration of the stationary wavelet transform (SWT) around the rising edge 
($i\sim6325$, 50ns per sample) of three 6.8 keVee CoGeNT events, each with a different value for the rise-time (see text).}
\end{figure}

\begin{figure}[!htbp]
\includegraphics[width=7cm]{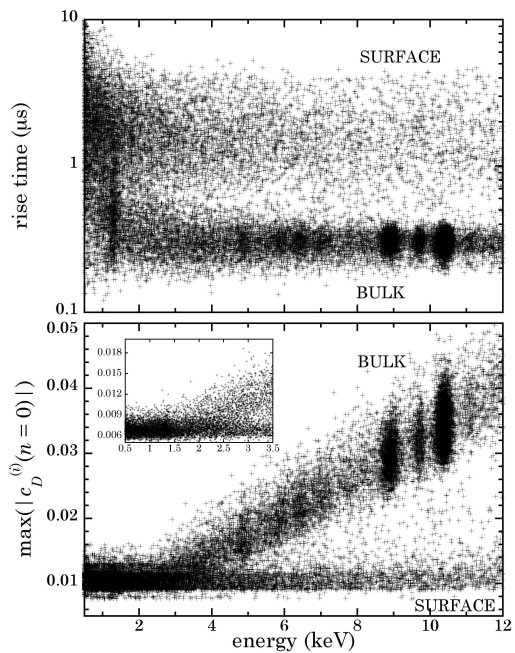}
\caption{{\it Top:} rise-time distributions for events from a 3.2 year CoGeNT exposure, showing two distinct populations down to threshold. {\it Bottom:} Maximum absolute value of the detail coefficients for these events (W$_{par}\!=\!$ max$(\mid\!c_{D}^{(i)}(n=0)\!\mid ^{2})/E^{2}$). The inset is a zoom-in of the same, extracted from the high-gain CoGeNT channel \protect\cite{longcogent}. W$_{par}$ does not correlate to rise-time at the lowest energies. It is instead a simple measure of how well the SWT can identify the presence of a rising edge in a preamplifier trace, which breaks down below $\sim$1 keVee for the present level of noise affecting both CoGeNT and MALBEK detectors, regardless of rise-time value. The merger of low-energy bulk events into the surface event distribution affecting MALBEK when using W$_{par}$ as a discriminator is made evident in this figure (see text).}
\end{figure}

The possible annual modulation observed in CoGeNT data below $\sim$2 keVee is a subtle effect, amounting to $\sim$5\% oscillations of the overall (bulk plus surface) signal rate \cite{cogentlast}. It is therefore of the utmost importance to accomplish the best possible separation between surface backgrounds and bulk signals, if a modulation of such a small magnitude stands a chance of being noticed \cite{note}. This is specially true when surface and bulk event rates are seen to be comparable, as is the case for both CoGeNT and MALBEK (Fig.\ 3), and when taking into consideration the statistics of small numbers involved in a modulation analysis for detectors this modest in mass. In view of the poor to non-existent surface/bulk discrimination ability accomplished within MALBEK in the energy region of interest below 1 keVee, it would be hard to assign any value to a MALBEK search for an annual modulation, if performed by means of a W$_{par}$ analysis. 

In conclusion, a careful inspection of the physical meaning of the parameter W$_{par}$ reveals a lack of correlation, even indirect, to preamplifier pulse rise-time t$_{10-90}$ below $\sim$1 keVee. With decreasing energy, W$_{par}$ is observed to rapidly collapse to values assigned to surface events, for all events. As a result, the majority of pulses in the energy region of interest for low-mass dark matter searches are erroneously identified as surface events in a recent MALBEK analysis \cite{malbek}. When applying aggressive W$_{par}$ cuts, this leads to an unwarranted decrease of the bulk event rate in MALBEK's low energy spectrum, and to an artificially enhanced sensitivity to low-mass dark matter particles. 

A similar concern can be expressed of a recent analysis of data from another PPC \cite{cdex}, where a dramatic reduction in low-energy background is claimed from use of a method involving extrapolations to low-energy from observations derived using high-energy gamma source calibrations. A straightforward validation of bulk event signal acceptance in the crucially relevant energy region near threshold, obtained through the use of an electronic pulser, is notoriously absent from \cite{cdex}. While PPCs offer the possibility to discern surface from bulk events, it is easy to overestimate or underestimate the magnitude of the irreducible spectrum next to detector threshold. The only solution to this riddle is a further hardware improvement to the PPC electronic noise. This has been accomplished in a next-generation CoGeNT detector, to be described in an upcoming publication.

An exponential excess of low-energy bulk events remains present in an upcoming analysis \cite{matt} of a high-statistics 3.4 year CoGeNT dataset \cite{cogentlast}. This excess is identifiable through a direct measurement of rise-time, the physically-relevant quantity when separating surface from bulk events in PPC detectors. The significance of this excess is however observed to depend on the choice of input to the analysis (use of electronic pulser calibrations, or instead of simulated pulses, when defining the characteristics of bulk events \cite{matt}). Such an excess is difficult to explain based on known backgrounds \cite{longcogent}, and is similar in magnitude and spectral shape to one recently found in the electron recoil band of superCDMS germanium detectors \cite{supercdms1}. The possibility of this spectral feature to originate in nuclear recoils seems now remote in view of the most recent superCDMS germanium results \cite{supercdms2}, but its origin and possible association to the low-significance modulation found in \cite{cogentlast} remain unknown. 

The shortcomings of the work presented in \cite{malbek} can be used to illustrate that the performance of a state-of-the-art low-noise detector needs to be matched by its data acquisition system, and other associated electronics. The choices made in this respect within MALBEK can limit the promising performance of a PPC-based MAJORANA Demonstrator as a low-mass dark matter detector \cite{Aal08}.

The author is indebted to N. Fields and P.S. Finnerty for useful exchanges.


\begin{thebibliography}{9}
\bibitem{note2}A dispersion in the value of t$_{10-90}$ is expected from the position-dependence of bulk events within the PPC crystal, which affects the charge collection time through variations in electric field intensity and path length to the collecting electrodes. When combined with the poor to non-existent surface/bulk discrimination power of W$_{par}$ at the lowest energies explored, a large fraction of actual  bulk events near threshold can be assigned significantly smaller (surface-like) values of W$_{par}$ than what is suggested by the use of a fixed t$_{10-90}$ electronic pulser.
\bibitem{note} While we have recently improved the choice of rise-time cuts leading to an irreducible bulk event spectrum for CoGeNT \cite{longcogent}, the algorithms used for rise-time measurement have remained unchanged since \cite{Aal11}, i.e., the optimal CoGeNT surface/bulk discrimination discussed here is due to the noise levels originally achieved by the detector and associated electronics, and not to any later improvements in the data analysis.
\end{thebibliography}
\end{document}